\documentclass[prl,
twocolumn,a4paper,nofootinbib,nobibnotes,showpacs,amsmath,amssymb,tbtags]{revtex4}
\usepackage{txfonts}
\renewenvironment{acknowledgements}{\section*{Acknowledgements}}{\par}
\usepackage[italian,french,german,british]{babel}
\usepackage{bm}
\usepackage{url}
\usepackage{hyperref}

\newcommand{\txgreek}{%
\newcommand{\rpi}{\piup}
\newcommand{\rde}{\mathrm{\partial}}
\newcommand{\rdelta}{\deltaup}
\newcommand{\rvarepsilon}{\varepsilonup}
\newcommand{\rmu}{\muup}
\newcommand{\rzeta}{\zetaup}
}
                          \txgreek
\newcommand{\pu}{\rpi}

\newcommand{\cond}{\mathbin{\mid}}
\newcommand{\with}{\colon}
\newcommand{\mult}{\times}
\newcommand{\defin}{\stackrel{\text{\tiny def}}{=}}
\renewcommand{\ge}{\geqslant}
\DeclareMathDelimiter{\lclose}{\mathopen}{operators}{"5B}{largesymbols}{"02}
\DeclareMathDelimiter{\rclose}{\mathclose}{operators}{"5D}{largesymbols}{"03}
\DeclareMathDelimiter{\lopen}{\mathopen}{operators}{"5D}{largesymbols}{"03}
\DeclareMathDelimiter{\ropen}{\mathclose}{operators}{"5B}{largesymbols}{"02}

\newcommand{\set}[1]{\{#1\}}
\newcommand{\pr}{P}
\newcommand{\sh}{H}
\newcommand{\shk}{K}
\newcommand{\prop}[1]{\textsf{``#1''}}
\newcommand{\uni}[1]{~\mathrm{#1}}
\newcommand{\degree}{^{\mathord{\circ}}}
\newcommand{\etc}{{etc.}}
\newcommand{\ie}{{i.e.}}
\newcommand{\eg}{{e.g.}}

\newcommand{\cf}{{cf.}}
\newcommand{\Cf}{{Cf.}}

\newcommand{\etal}{{et al.}}

\newcommand{\sect}{{\S}}
\newcommand{\sects}{{\S\S}}
\newcommand{\cm}{\uni{cm}}
\newcommand{\bquoteq}{\text{`` }}
\newcommand{\equoteq}{\text{ ''}}
\newcommand{\zwi}{T}
\newcommand{\zhe}{R}
\newcommand{\zwia}{2\cm}
\newcommand{\zhea}{3\cm}
\newcommand{\ztoa}{5\cm}
\newcommand{\zwib}{20\cm}
\newcommand{\zheb}{30\cm}
\newcommand{\ztob}{50\cm}
\newcommand{\bit}{\uni{bit}}
\newcommand{\shb}[1]{\sh[#1]}
\newcommand{\shp}[1]{\sh(#1)}
\newcommand{\shh}[1]{\sh\bigl(#1\bigr)}
\newcommand{\zA}{a}
\newcommand{\zAo}{\zA_{\text{out}}}
\newcommand{\zAn}{\zA_{\text{not-out}}}
\newcommand{\zAA}{A}
\newcommand{\zAAA}{\set{\zA_i \with i=1,\dotsc, n_\zA}}
\newcommand{\zB}{b}
\newcommand{\zBo}{\zB_{\text{out}}}
\newcommand{\zBn}{\zB_{\text{not-out}}}
\newcommand{\zBB}{B}

\newcommand{\zAB}{\zAA\land\zBB}
\newcommand{\zABB}{\set{\zA_i\land\zB_j \with i=1,\dotsc, n_\zA;j=1,\dotsc, n_\zB}}
\newcommand{\zX}{a}
\newcommand{\zXu}{\zX_{\text{up}}}
\newcommand{\zXd}{\zX_{\text{down}}}
\newcommand{\zXX}{A}

\newcommand{\zY}{b}
\newcommand{\zYu}{\zY_{\text{up}}}
\newcommand{\zYd}{\zY_{\text{down}}}
\newcommand{\zYY}{B}

\newcommand{\zL}{l}
\newcommand{\zLurn}{l_{\text{u}}}
\newcommand{\zLone}{l_{\text{one}}}
\newcommand{\zLL}{l_{\text{two}}}

\newcommand{\zN}{n}
\newcommand{\zNL}{n_{\text{inv}}}
\newcommand{\zM}{m}
\newcommand{\zML}{m_{\text{inv}}}
\newcommand{\zMM}{q}
\newcommand{\zMML}{q_{\text{inv}}}
\newcommand{\zK}{k}
\newcommand{\zKL}{k_{\text{inv}}}
\newcommand{\za}{p}
\newcommand{\zauu}{p_{ab}}
\newcommand{\zaud}{p_{a\Bar{b}}}
\newcommand{\zadu}{p_{\Bar{a}b}}
\newcommand{\zadd}{p_{\Bar{a}\Bar{b}}}
\newcommand{\zXb}{\zX_{\text{black}}}
\newcommand{\zXw}{\zX_{\text{white}}}

\newcommand{\zYp}{\zY_{\text{plastic}}}
\newcommand{\zYw}{\zY_{\text{wood}}}
\allowdisplaybreaks[1]

\begin{document}
\bibliographystyle{apsrevmana}
\title{Consistency of the Shannon entropy in quantum experiments}

\author{Piero\;G.\;Luca Mana}

\email{mana@imit.kth.se}

\affiliation{Institutionen f\"or Mikroelektronik och
  Informationsteknik,
Kungliga Tekniska H\"ogskolan,
Isafjordsgatan 22, SE-164\,40 Kista, Sweden}

\date{16 July 2004}

\begin{abstract}
The consistency of the Shannon entropy, when applied to outcomes of
quantum experiments, is analysed. It is shown that the Shannon
entropy is fully consistent and its properties are never violated in
quantum settings, but attention must be paid to logical and
experimental contexts. This last remark is shown to apply regardless
of the quantum or classical nature of the experiments.
\end{abstract}


\maketitle

\section{\label{sec:intro}Introduction}

Since its introduction in 1948, the Shannon
entropy~\cite{shannon1948} has played a central r\^ole in all
branches of communication
theory, 
where it allows a precise and operational definition of useful
concepts like transmission rate and channel capacity, and has found
many important applications in other branches of science as
well~\cite{tribus1979,zellner1980}.  Shannon introduced it to
quantify the amount of ``uncertainty'' or ``choice'' present in a
probability distribution for a set of messages or symbols. In this
regard, his entropy possesses some mathematical
properties~\cite{shannon1948,aczeletal1975,wehrl1978,wehrl1977,cox1961,aczel1984,ochs1976}
that correspond to intuitive properties expected from a measure of
uncertainty.

In a recent paper~\cite{brukneretal2001}, however, Brukner and
Zeilinger analyse the application of the Shannon entropy to the
statistical outcomes of two simple quantum-mechanical
thought-experiments and of a non-quantum-mechanical, or `classical',
one, and conclude that two particular properties of the Shannon
entropy are violated in quantum experiments while holding in
classical ones.

This conclusion 
poses serious limitations when we wish to use the Shannon entropy,
because of its particular properties, in quantum-mechanical problems.
It would be therefore reasonable to ascertain in which sense or to
what extent (if any) the claimed violations occur.

Deep criticisms of the conclusions of
Ref.~\citep{brukneretal2001} have already appeared in the
literature, mainly Hall's~\cite{hall2000} and
Timpson's~\cite{timpson2003,timpson2003b}. These
interesting criticisms are mainly of a theoretical (and
partly philosophical) nature, and so do not explain
whether or how the concrete and interesting
thought-experiments given in Ref.~\citep{brukneretal2001}
really represent a violation of the Shannon entropy's
properties.

The purpose of the present paper is more pragmatic: to re-analyse
the experiments presented in Ref.~\citep{brukneretal2001} and to
show \emph{by explicit calculation} that \emph{they present in fact
no violation of any property of the Shannon entropy}.

We shall see that the seeming ``violations'' appear only
because the authors of Ref.~\citep{brukneretal2001}
tacitly shift between different contexts (different
experimental arrangements), thus misapplying Shannon's
formulae. In order to explain what is meant here by
`context', and to make the last point clearer, let us
consider the following example:\footnote{A less informal
example is presented in the published version of this
paper, Phys.\ Rev.~A \textbf{69}, 062108 (2004).}

In another Wonderland, Alice is acquainted with two Cheshire Cats: a
Frumious one and a very Mome one\nocite{Carroll1865,Carroll1871}. The
Frumious Cheshire Cat always appears with its tail ($\zwi$) first, and then
with the rest ($\zhe$) of its body (head included), and in these apparitions
Alice always observes first only that $\zwi= \zwia$, and then that
$\zhe=\zhea$, hence
\begin{equation}
\label{eq:hearts}
\bquoteq\zwi+\zhe= \ztoa \equoteq.
\end{equation}
The Mome Cheshire Cat, instead, always appear with its
rest ($\zhe$) first, and then comes the tail ($\zwi$), so
that, when meeting it, Alice can first observe that $\zhe=
\zheb$, and then that $\zwi=\zwib$; hence
\begin{equation}
\label{eq:diamonds}
\bquoteq\zhe+\zwi= \ztob\equoteq.
\end{equation}
Looking at Eqs.~\eqref{eq:hearts} and~\eqref{eq:diamonds}, Alice
concludes:
\begin{equation}
\label{eq:wrongresal}
\bquoteq\zwi+\zhe\neq \zhe+\zwi\equoteq,
\end{equation}
or: ``the commutative property of addition is violated for Cheshire
Cats''.

Is Alice's conclusion about the commutative property correct? No.
First of all, we see that the content of Eq.~\eqref{eq:hearts} is
really just ``$\zwia+\zhea=\zhea+\zwia=\ztoa$'', and analogously for
Eq.~\eqref{eq:diamonds}, hence the commutative property is in fact
clearly satisfied in those equations.

On the other hand, what Alice means by Eq.~\eqref{eq:wrongresal} is
just ``$\zwia+\zhea\neq \zheb+\zwib$'', so that this equation has little to
do with the commutative property of addition. In fact
Eq.~\eqref{eq:wrongresal}, as it is written, makes simply no sense,
because the various mathematical symbols therein are inconsistently
used: they have different values on the right- than on the left-hand
side.

This comes about because Alice tacitly shifts from a given
context, where the symbols `$\zwi$' and `$\zhe$' refer to
the Frumious Cheshire Cat and have some values, to a
different context, where the same symbols refer to the
Mome Cheshire Cat and have different values --- but the
commutative property of addition is not meant to apply
that way. Related to this is the fact that Alice seems to
attribute a \emph{temporal} meaning to the \emph{order} in
which the summands appear in the
additions~\eqref{eq:hearts}--\eqref{eq:wrongresal}, but we
know that the order of terms in an addition has no
temporal meaning. If Alice wishes to express temporal
details, she can do so by, \eg, appending subscripts to
the symbols, like $\zwi_{t'}$, $\zhe_{t''}$, \etc; in any
case she cannot burden the addition operation with
meanings that the latter does not have.

Hence, Alice's incorrect conclusion stems from her using,
inconsistently, the \emph{same} symbols $\zwi$ and $\zhe$
for two different and incompatible logical and
experimental contexts. She could have used, \eg,
$\zwi_{\text{Mome},t'}$, $\zhe_{\text{Frum},t''}$, \etc, instead,
writing her findings as
\begin{multline}
\label{eq:aliceneweq}
\zwi_{\text{Mome},t'}+\zhe_{\text{Mome},t''}=
\zhe_{\text{Mome},t''}+\zwi_{\text{Mome},t'}\\
{}\neq\zwi_{\text{Frum},t''}+\zhe_{\text{Frum},t'}=
\zhe_{\text{Frum},t'}+\zwi_{\text{Frum},t''},
\end{multline}
where all symbols are consistently and informatively used
and it is clear that no violations of any mathematical
property occur.

\emph{In nuce}, an analogous \emph{faux pas} is made in
Ref.~\citep{brukneretal2001}, although less apparent,
because there instead of the addition we have more complex
mathematical functions --- probabilities and entropies
---, and quantum experiments take the place of the
example's Cheshire-Cats experiments. In the mentioned
paper, the authors compare probabilities and Shannon
entropies pertaining to different contexts (different
experimental arrangements); thus their equations do not
pertain to the properties they meant to discuss. To my
knowledge, this simple point has hitherto never been
noticed in the literature, and will be shown here by
explicit calculation.

We shall also investigate how the peculiar reasoning in
Ref.~\citep{brukneretal2001} comes about, and find that it stems
from an improper temporal interpretation of the logical symbols
`$\cond$' and `$\land$'. This point has not been noticed in the
literature either.\footnote{In fact, even
Timpson~\cite[p.~15]{timpson2003} seems to attach an improper
temporal sense to the logical concept of `joint probability'.}

It is also clear that such kinds of \emph{seeming}
``violations'', arising through mathematical and logical
misapplication, can then easily be made to appear not only
in quantum experiments, but in classical ones as well, and
this will also be shown by simple examples.

For these purposes, some notation and definitions for probabilities
and the Shannon entropy will first be introduced, where the r\^ole of
the context --- \eg, the given experimental arrangement --- is emphasized.
Then, after a discussion of the two questioned properties of
the entropy, the experiments of Ref.~\citep{brukneretal2001}
will be presented and re-analysed. Finally, two ``counter-experiments''
will be presented.

\section{Definitions and notation\label{sec:notat}}
We shall work with probability theory as extended
logic~\citep{jeffreys1939,cox1961,jaynes2003}, so we denote by
$\pr(\zA\cond \zL)$ the probability of the proposition~$\zA$ given
the prior knowledge~$\zL$. This prior knowledge, also called (prior)
data or prior information, represents the \emph{context} the
proposition refers to. It is useful to write explicitly the prior
knowledge --- a usage advocated by Keynes~\cite{keynes1921},
Jeffreys~\cite{jeffreys1931,jeffreys1939},
Koopman~\cite{koopman1940a,koopman1940b,koopman1957},
Cox~\cite{cox1946,cox1961,cox1979}, Good~\cite{good1950}, and
Jaynes~\cite{jaynes2003} --- 
because the meaning and the probability of a proposition always
depend on a given context. Compare, \eg,\ 
$\pr({}$\prop{Tomorrow it will snow}${}\cond{}$\prop{We are in Stockholm
and it is winter}${})$ and $\pr({}$\prop{Tomorrow it will
snow}${}\cond{}$\prop{We are in Rome and it is summer}${})$.%
\footnote{The relevance of prior knowledge is nicely
expressed in the following brief dialogue, which I could
not resist quoting, between Holmes and
Watson~\citep{doyle1891}, taking place after the two
listened to a client's statement: \protect\begin{quote}
\mbox{\hspace{3ex}}``I think that I shall have a whisky
and soda and a cigar after all this cross-questioning. I
had formed my conclusions
as to the case before our client came into the room.''\\
\mbox{\hspace{3ex}}``My dear Holmes!''\\
\mbox{\hspace{3ex}}``[\ldots]\ My whole examination served
to turn my conjecture into a certainty. Circumstantial
evidence is occasionally very convincing, as when you find
a trout in the milk, to quote
Thoreau's example.''\\
\mbox{\hspace{3ex}}``But I have heard all that you have heard.''\\
\mbox{\hspace{3ex}}``Without, however, the knowledge of
pre-existing cases which serves me so well.''
\protect\end{quote}} Even more, the probability of a
proposition can be undefined in a given context (\ie, that
proposition is meaningless in that context): consider,
\eg, $\pr({}$\prop{My daughter's name is
Kristina}${}\cond{}$\prop{I have no children}${})$. Common
usage tends to omit the context and writes $\pr(\zA)$
instead of $\pr(\zA\cond \zL)$, but this usage can in some
cases --- especially when the context is \emph{variable},
\ie, several different contexts are considered in a
discussion --- lead to ambiguities, as will be shown.

In particular, a proposition can represent an outcome of an
experiment or observation that was, is, or will be, performed. The
context in this case consists of all the details of the experimental
arrangement and of the theoretical model of the latter which are
necessary to assign a probability to that outcome. This is true for
both ``classical'' and ``quantum'' experiments.

The importance of the context is also evident when considering a set
of propositions
\begin{equation}
\zAA\defin \zAAA
\end{equation}
such that, in the context~$\zL$, they are mutually exclusive and
exhaustive (\ie, one and only one of them is true). Indeed, it is
clear that the propositions may be mutually exclusive and exhaustive
in some context while not being such in another. For example,
the propositions $\{{}$\prop{A red ball is drawn at the first
draw}${},{}$\prop{A wooden ball is drawn at the first draw}${}\}$ are
mutually exclusive and exhaustive in the context \prop{Balls are
drawn from an urn containing one red plastic ball and one blue wooden
ball}, but are neither mutually exclusive nor exhaustive in the
context \prop{Balls are drawn from an urn containing one red
wooden ball, one red plastic ball, and one blue plastic ball}, and
make no sense in the context \prop{Cards are drawn from a deck}.

In the following, propositions which are mutually exclusive and
exhaustive in a given context will be called \emph{alternatives}, and
their set a \emph{set of alternatives}.\footnote{Another name for
such a set could be `event' (\cf\ Tribus~\cite{tribus1961}); Cox
uses the terms `(inductive) system'~\cite{cox1961} and (with
a slightly more general meaning)`question'~\cite{cox1979}.}
Such a set may represent the possible, mutually exclusive outcomes in
a given experiment.

The probability distribution for a set of alternatives $\zAAA$ in the
context~$\zL$ will be denoted by
\begin{gather}
\pr(\zAA\cond\zL)\defin\set{\pr(\zA_i\cond\zL)\with  i=1,\dotsc,n_\zA},\\
\intertext{and has, of course, the usual properties}
\pr(\zA_i\land\zA_{i'}\cond \zL)=0\qquad\text{for $i\neq i'$},\\
\sum_i\pr(\zA_i\cond \zL)=1,
\end{gather}
where the
\emph{conjunction}~\cite{lewisetal1932,copi1953,hamilton1978}
(sometimes also called `logical product') of $\zX$ and $\zY$ is
denoted by $\zX\land\zY$.

The Shannon entropy~\cite{shannon1948} is a function of the
probability distribution for a given set of alternatives~$\zAA$
\emph{in a given context~$\zL$}. It is defined as
\begin{equation}
\shb{\pr(\zAA\cond \zL)} \defin  -\shk\sum_i \pr(\zA_i\cond \zL)
\ln  \pr(\zA_i\cond \zL),
\quad\text{$\shk>0$},
\label{eq:entropy}\end{equation}
with the usual conventions for the units in which it is expressed and
the positive constant~$\shk$, and the limiting procedure when
vanishing probabilities are
present.\footnote{Cox~\cite{cox1961,cox1979} gives a generalisation
of Shannon's formula which is valid for a general set of not
necessarily mutually exclusive propositions.} The Shannon entropy can
be considered to quantify the amount of ``uncertainty'' associated
with a probability distribution.\footnote{A very small sample of the
literature on Shannon's entropy is represented by the work of
Kullback~\cite{kullback1959}, R{\'e}nyi~\cite{renyi1970}, Acz{\'e}l
\etal~\cite{aczeletal1975,aczel1984}, Wehrl~\cite{wehrl1978}, Shore
\etal~\cite{shoreetal1980}, Jaynes~\cite{jaynes2003}; see also
Forte~\cite{forte1984}.}
 
Given two sets of alternatives $\zAA$ and $\zBB$ in the
context~$\zL$, the set of all conjunctions of the propositions
$\set{\zA_i}$ with the propositions $\set{\zB_j}$,
\begin{equation}
\zAB\defin \zABB,
\end{equation}
is also a set of alternatives in~$\zL$, as follows from
the rules of probability theory; it will be called a
\emph{composite set}.\footnote{I prefer the term
`composite' to the term `joint', as the latter seems to
bring about improper temporal meanings.} Thus the
composite probability distribution
\begin{multline}
\pr(\zAB\cond\zL)\defin\{\pr(\zA_i\land\zB_j\cond\zL)\with
  i=1,\dotsc,n_\zA; j=1,\dotsc,n_\zB\}
\end{multline}
satisfies the properties
\begin{gather}
\pr\bigl((\zA_i\land\zB_j) \land
(\zA_{i'}\land\zB_{j'})\cond \zL\bigr)= 0\quad\text{for $i\neq i'$ or $j\neq j'$},\\
\sum_{ij}\pr(\zA_i\land\zB_j\cond \zL) = 1.
\end{gather}

The probability distributions for the sets of alternatives $\zAA$,
$\zBB$, and $\zAB$ in the context $\zL$ are related by the
following standard probability rules (see \eg\ 
Jaynes~\cite[Ch.~2]{jaynes2003},
Jeffreys~\cite[Ch.~1]{jeffreys1939}\cite[Ch.~2]{jeffreys1931},
Cox~\cite[Ch.~1]{cox1961})
\begin{subequations}\label{eq:probprop}
\begin{gather}
\begin{aligned}
\pr(\zA_i\cond\zL)&=\sum_j \pr(\zA_i\land\zB_j\cond\zL),\\
\pr(\zB_j\cond\zL)&=\sum_i \pr(\zA_i\land\zB_j\cond\zL)
\end{aligned}
\label{eq:marginalprob}\\
\intertext{(marginal probabilities), and}
\begin{aligned}
\pr(\zA_i\land\zB_j\cond\zL)&=
\pr(\zA_i\cond\zB_j\land\zL)\; \pr(\zB_j\cond\zL)\\
& =\pr(\zB_j\cond\zA_i\land\zL)\; \pr(\zA_i\cond\zL)
\end{aligned}
\label{eq:productrule}\\
\intertext{(product rule), from which}
\begin{aligned}
&\pr(\zA_i\cond\zB_j\land\zL)=\frac{\pr(\zA_i\land\zB_j\cond\zL)}{\pr(\zB_j\cond\zL)}\quad\text{if
$\pr(\zB_j\cond\zL)\neq0$,}\\
&\pr(\zB_j\cond\zA_i\land\zL)=\frac{\pr(\zA_i\land\zB_j\cond\zL)}{\pr(\zA_i\cond\zL)}\quad\text{if
$\pr(\zA_i\cond\zL)\neq0$}\\
\end{aligned}
\label{eq:bayesrule}
\end{gather}
\end{subequations}
(Bayes' rule) follow.

All the probability distributions for the sets of alternatives $\zAA$, $\zBB$,
and $\zAB$ have associated Shannon entropies. It is
also possible to define the \emph{conditional entropy} of the
distribution for~$\zBB$ relative to the distribution for~$\zAA$, as
follows:
\begin{multline}\label{eq:shancond}
  \shb{\pr(\zBB\cond \zAA\land\zL)} \defin \sum_{i} \pr(\zA_i\cond
  \zL) \shb{\pr(\zBB\cond \zA_i\land\zL)}\\
 =-\shk\sum_{i} \pr(\zA_i\cond\zL)  \sum_j \pr(\zB_j\cond \zA_i\land\zL) \ln \pr(\zB_j\cond
  \zA_i\land\zL).
\end{multline}
An analogous definition is given for $\shb{\pr(\zAA\cond
  \zBB\land\zL)}$.

\section{\label{sec:sec}Properties of the Shannon entropy and quantum
experiments}

The various Shannon entropies presented in the previous section
possess several mathematical
properties~\cite{shannon1948,aczeletal1975,wehrl1978,wehrl1977,cox1961,aczel1984,ochs1976}
which have a fairly intuitive meaning when the Shannon entropy is
interpreted as a measure of ``uncertainty''. We shall focus our
attention on two of these properties which, it is
claimed~\cite{brukneretal2001}, are violated in quantum
experiments.

Consider the probability distribution for a composite set of
alternatives~$\zAA\land\zBB$ in a context~$\zL$. The entropy
$\shb{\pr(\zBB\cond \zL)}$ and the conditional entropy
$\shb{\pr(\zBB\cond \zAA\land\zL)}$ satisfy
\begin{subequations}\label{eq:prop}
\begin{equation}
 \shb{\pr(\zBB\cond \zL)} \ge
  \shb{\pr(\zBB\cond \zAA\land\zL)}.\label{eq:prop1}
\end{equation}
This property (also called \emph{concavity} since it arises from the
concavity of the function $-x\ln x$), intuitively states that the
uncertainty of the probability distribution for $\zBB$ in the context
$\zL$ decreases or remains the same, \emph{on average}, when the
context is ``updated'' because one of the $\set{\zA_i}$ is known to
be true. An \emph{intuitive} picture can be given as follows. Imagine
that someone performs an experiment (represented by $\zL$) consisting
in two observations (represented by the sets $\zAA$ and $\zBB$), and
writes the results of the experiment in an observation record, say,
on a piece of paper denoted by `$\zL$', under the headings `$\zAA$'
and `$\zBB$'. Now suppose that we know all the details of the
experiment except for the outcomes: we have not yet taken a look at
the record and are uncertain about what result is written under
`$\zBB$'. If we now read the result written under `$\zAA$', our
uncertainty about the result under `$\zBB$' will decrease or remain
the same on average (\ie, in most, though \emph{not all}, cases),
since the former result can give us some clues about the latter.

The second property (\emph{strong additivity}~\cite{aczeletal1975})
reads
\begin{equation}
\begin{split}
\shb{\pr(\zAB\cond\zL)} &=
\shb{\pr(\zAA\cond\zL)} + \shb{\pr(\zBB\cond \zAA\land\zL)}\\
&= \shb{\pr(\zBB\cond\zL)} + \shb{\pr(\zAA\cond \zBB\land\zL)},
\end{split}\label{eq:prop2}
\end{equation}
\end{subequations}
and its intuitive meaning is that the uncertainty of the probability
distribution for the composite set of alternatives $\zAB$ is given by
the sum of the uncertainty of the probability distribution for $\zAA$
and the average uncertainty of the updated probability distribution
for $\zBB$ if one of the $\set{\zA_i}$ is known to be true, or
\emph{vice versa}. Using the intuitive picture already proposed, we
are initially uncertain about both outcomes written on the record
`$\zL$'. This uncertainty will first decrease as we read the outcome
for `$\zAA$', and then disappear completely when we read the outcome
for `$\zBB$' (given that we do not forget what we have read under
`$\zAA$'). Equivalently, the total uncertainty will first decrease
and then disappear as we read first the outcome under `$\zBB$' and
then the one under `$\zAA$'.

Three remarks are appropriate here. The first is that \emph{the above
are mathematical, not physical, properties}: they are not
experimentally observed regularities, but rather follow mathematically,
through the properties of basic arithmetical functions like the
logarithm, once we put some numbers (the probabilities
$\pr(\zA_1\cond \zL)$, $\pr(\zB_2\cond \zL)$, $\pr(\zB_2\cond
\zA_1\land\zL)$, \etc)\ into the formulae~\eqref{eq:entropy}
and~\eqref{eq:shancond}. In particular, \emph{they can be neither
proven nor disproved by experiment}, but only correctly or
incorrectly applied. It is for this reason that one suspects that the
seeming violations found in Ref.~\citep{brukneretal2001} are
only caused by an incorrect application of the
formulae~\eqref{eq:prop}; we shall see that this is the case, because
in the mentioned paper some probabilities are used in the right-hand
sides of Eqs.~\eqref{eq:prop}, and (numerically) \emph{different}
probabilities in the left-hand sides.

This leads us to the second remark: \emph{the above properties hold
when all the probabilities in question refer to the same context};
otherwise, they are not guaranteed to hold. This is because if a
specific set of alternatives, in two different contexts, has two
different sets of probabilities, then it will yield, in general, two
different entropies as well. Hence we cannot expect the
formulae~\eqref{eq:prop} to hold if we use, in the left-hand sides,
the probabilities relative to a given experiment, and, in the
right-hand sides, those relative to a different
experiment. This remark amounts simply to saying that the
expression `$x+y=y+x$' holds only if we give $x$ and $y$ the same value
on both sides (\cf\ Alice's example in the Introduction). The
(somewhat pedantic) notation $\shb{\pr(\zAA\cond \zL)}$ and
$\shb{\pr(\zBB\cond \zAA\land\zL)}$, instead of the more common
$\shp{\zAA}$ and $\shp{\zBB\cond\zAA}$, is used here to stress this
point.

The third remark (\cf\ also Koopman~\cite{koopman1957}) is that an
expression like `$\zBB\cond\zAA$' does not imply that ``the
observation represented by $\zAA$ is performed \emph{before} the one
represented by $\zBB$''; nor does `$\zAA\land\zBB$' mean that the two
observations are carried out ``simultaneously''. The temporal order
of the observations is formally contained in the context $\zL$,
and \emph{the symbols `\/$\cond$\/' and `\/$\land$\/' have a logical, not
temporal, meaning}. One must be careful not to confuse logical
concepts with mathematical objects or physical procedures, even when
there may be some kind of relationship amongst them; this kind of
confusion is strictly related to what Jaynes called ``the mind
projection fallacy''~\cite{jaynes1989,jaynes1990,jaynes2003}.

Consider, \eg, the following situation: an urn contains a red, a
green, and a yellow ball, and we draw two balls without replacement
(this is our context $\zLurn$). Consider the propositions
$\zA\defin{}$\prop{Red at the first draw}, and $\zB\defin{}$\prop{Green
at the second draw}. Now, suppose that we also know that the second draw
yields `green'; what is the probability that the first draw yields
`red'? The answer, to be found in any probability-theory textbook, is
$\pr(\zA\cond\zB\land\zLurn)= \pr(\zA\land\zB\cond\zLurn)/\pr(\zB\cond\zLurn)=
(1/3)\times (1/2)/(1/3)= 1/2$. We see that \emph{the expression
`$\zA\cond\zB$' does not mean that $\zB$ precedes $\zA$} (nor does it
mean that $\zB$ ``causes'' $\zA$): it just expresses a \emph{logical}
connexion. We can also ask: what is the probability of obtaining
\emph{first} `red' and \emph{then} `green'? The answer is again standard:
$\pr(\zA\land\zB\cond\zLurn)= 1/6$. We see that \emph{the expression
`$\zA\land\zB$' does not mean that $\zA$ and $\zB$ happen
simultaneously}: it just expresses a \emph{logical}
connexion.\footnote{We can change the example by considering,
instead, the order in which three identically prepared radioactive
nuclei, in three different laboratories, will decay. The analysis and
the conclusion will nevertheless be the same. `Quantumness' plays no
r\^ole here.}

Note that Shannon himself, in his paper, does not attribute
any temporal meaning to the conditional or joint entropies, but only
the proper logical one. He uses, \eg~\cite[\sects11 and
12]{shannon1948}, the expressions $\shb{\pr(X\cond Y\land\zL)}$ and
$\shb{\pr(X\land Y\cond\zL)}$ (which in his notation are `$H_{y}(x)$'
and `$H(x,y)$'~\cite[\sect6]{shannon1948}) where `$Y$' represents
the \emph{received} signal, while `$X$' represents the signal
\emph{source} --- \ie, where $X$ actually \emph{precedes}
$Y$.\footnote{From Shannon's paper: ``First there is the entropy
$H(x)$ of the source or of the input to the channel [\ldots]. The
entropy of the output of the channel, \ie, the received signal, will
be denoted by $H(y)$. [\ldots] The joint entropy of input and output
will be $H(x,y)$. Finally there are two conditional entropies
$H_x(y)$ and $H_y(x)$, the entropy of the output when the input is
known and conversely. Among these quantities we have the relations
$H(x,y)=H(x)+H_x(y)=H(y)+H_y(x)$''~\cite[\sect11]{shannon1948}.} We
may also remark that Shannon's entropy formulae for channel
characteristics are not directly dependent on the actual physical
details of the system(s) constituting the channel, which can be
classical, quantum, or even made of pegasi and unifauns. Only their
statistical properties are directly relevant.

Hence, the conjunction $\zA_i\land\zB_j\equiv\zB_j\land\zA_i$
(and $\zBB\land\zAA\equiv\zAA\land\zBB$) is commutative even if the
matrix product of two operators which can in some way be associated
with these propositions is not.

We shall see that these remarks have indeed a bearing on
the analysis presented in Ref.~\citep{brukneretal2001},
where the validity of the formulae~\eqref{eq:prop} was
analysed in two quantum experiments and a classical one,
and it was found that, \emph{seemingly},
Eqs.~\eqref{eq:prop} did not hold in the quantum case,
while they still held in the classical one. These
experiments will be now presented using first their
original notation of Ref.~\citep{brukneretal2001}
(indicated by the presence of quotation marks), which
makes no reference to the context, and then re-analysed
using the expanded notation introduced above.

\subsection{First quantum experiment\label{sec:first}}
The first quantum experiment~\cite[Fig.~4]{brukneretal2001}
runs as follows. Suppose we send a vertically polarised photon
through a horizontal polarisation filter. The set of
alternatives~$\zBB\defin\set{\zBo,\zBn}$ refers to the photon's
coming out of the filter, with $\zBo\defin{}$\prop{The photon comes
out of the horizontal filter}, $\zBn\defin{}$\prop{The photon does
not come out of the horizontal filter}. Since we are sure about
$\zBn$, \ie, ``$\pr(\zBn)=1$'', we have that
\begin{equation}\label{eq:shb}
\bquoteq\shp{\zBB}=0\equoteq.  
\end{equation}
Then we insert a diagonal ($45\degree$) filter before the horizontal
one; the set of alternatives~$\zAA$ refers to the photon's coming out
of the diagonal filter, with $\zAo$ and $\zAn$ defined analogously.
Now, if we know that the photon has come out of the diagonal filter, we
are no longer sure that it will not get through the horizontal
filter, and so the uncertainty on $\zBB$ is increased by knowledge of
$\zAA$:
\begin{equation}
  \label{eq:sha}
\bquoteq\shp{\zBB\cond\zAA}>0\equoteq.
\end{equation}
Thus we find
\begin{equation}\label{inc1}
\bquoteq0=\shp{\zBB} < \shp{\zBB\cond\zAA}\equoteq\quad\text{(seemingly)}
\end{equation}
and property~\eqref{eq:prop1} is seemingly violated.

Only seemingly, though. The fact that a diagonal filter
was considered in the reasoning leading to
Eq.~\eqref{eq:sha} but \emph{not} in the reasoning leading
to Eq.~\eqref{eq:shb}, makes us doubt whether the
comparison of the entropies~``$\shp{\zBB}$''
and~``$\shp{\zBB\cond\zAA}$'' really corresponds to
Eq.~\eqref{eq:prop1}: these entropies, in fact, apparently
refer to two different experimental arrangements --- \ie, to
two different contexts. It becomes evident that this is
indeed the case if we proceed by calculating numerically
and explicitly \emph{all} the probabilities first, and
then the entropies, keeping the context in view.

Let us consider again the first experimental set-up, which will be
denoted by~$\zLone$, where \emph{only one}, horizontal, filter is
present. For the set~$\zBB$, we have of course that, using
our notation,
\begin{equation}\label{eq:prb1}
\pr(\zBo\cond \zLone)=0,\qquad\pr(\zBn\cond \zLone)=1,  
\end{equation}
so that the Shannon entropy is
\begin{equation}\label{eq:shb1}
\shb{\pr(\zBB\cond \zLone)}=\shp{0,1}=0\bit.
\end{equation}
At this point, asking for the probabilities for the set of
alternatives~$\zAA$, we realise that the propositions
$\zAo\defin{}$\prop{The photon comes out of the diagonal
filter} and $\zAn\defin{}$\prop{The photon does not come
out of the diagonal filter} make no sense here, since no
diagonal filter is present; consequently, there exist
\emph{no} entropies like $\shb{\pr(\zAA\cond \zLone)}$ or
$\shb{\pr(\zBB\cond \zAA\land \zLone)}$. So in this
experimental set-up it is not even meaningful to consider
the property~\eqref{eq:prop1}.

When we insert a diagonal filter before the horizontal one, we have a
new, \emph{different} experimental arrangement, which will be denoted
by~$\zLL$. In this new set-up it does make sense to speak of both
sets $\zAA$ and $\zBB$.\footnote{In the original formulation of the
example~\cite{brukneretal2001}, the authors denote ``by $\zAA$ and
$\zBB$ the properties of the photon to have polarization at
$+45\degree$ and horizontal polarization, respectively''; so that
$\zAA$ should perhaps be defined as $\{{}$\prop{The photon has diagonal
($45\degree$) polarisation}${},{}$\prop{The photon has no diagonal
($45\degree$) polarisation}${}\}$, and $\zBB$ analogously.  However,
there are problems with these propositions. If the photon is absorbed
by the diagonal filter, then it makes no sense to say that the photon
has no diagonal polarisation, since the photon is no longer present
(note that this problem has nothing to do with the non-existence of
properties of a system before an observation: the point is that, if
no photon is present, then it does not make sense to speak about its
properties anyway).  For this reason the alternative propositions
$\zAo$, $\zBn$, \etc, have been used here; however, this has not
affected the point of the experiment in
Ref.~\citep{brukneretal2001} --- namely, the seeming violation of
property~\eqref{eq:prop1}.} Quantum mechanics yields the following
probabilities:
\begin{gather}
\pr(\zAo\cond \zLL)=\tfrac{1}{2},\quad\pr(\zAn\cond
 \zLL)=\tfrac{1}{2},
\label{eq:pra2}\\     
\begin{aligned}
&\pr(\zBo\cond \zAo\land\zLL)=\tfrac{1}{2},\\
&\pr(\zBn\cond \zAo\land\zLL)=\tfrac{1}{2},
\end{aligned}
\label{eq:prbao}\\
\begin{aligned}
&\pr(\zBo\cond \zAn\land\zLL)=0,\\
&\pr(\zBn\cond \zAn\land\zLL)=1.
\end{aligned}\label{eq:prban}
\intertext{whence, by the product rule, it follows that} 
\pr(\zBo\cond \zLL)=\tfrac{1}{4},\quad\pr(\zBn\cond \zLL)=\tfrac{3}{4}.
  \label{eq:prb2}
\end{gather}
At this point we notice that the probabilities for the
set~$\zBB$, Eqs.~\eqref{eq:prb2}, \emph{differ
numerically} from those calculated in the previous set-up,
Eqs.~\eqref{eq:prb1}. This difference forces us to take
note of the difference of the set-ups~$\zLone$ and~$\zLL$;
if we ignored this, inconsistencies would arise already
for the probabilities, even \emph{before} computing any
entropy.

The Shannon entropy for the probability distribution for the
set~$\zBB$ appropriate in this context is readily calculated:
\begin{equation}
\shb{\pr(\zBB\cond\zLL)}=
\shh{\tfrac{1}{4},\tfrac{3}{4}}\approx0.81\bit,\\
\end{equation}
and we see that, in fact, it differs from the one in the first
set-up: thus, \emph{in the first quantum example of
Ref.~\textnormal{\citep{brukneretal2001}} the expression ``$\shp{\zBB}$'' is
inconsistently and ambiguously used}.

The conditional entropy relative to~$\zAA$ is:
\begin{equation}
\shb{\pr(\zBB\cond \zAA\land\zLL)}=0.5\bit,
\end{equation}
and, as a consequence,
\begin{equation}\label{eq:con1}
0.81\bit\approx\shb{\pr(\zBB\cond \zLL)}
\ge\shb{\pr(\zBB\cond \zAA\land\zLL)}=0.5\bit,
\end{equation}
in accord with the property~\eqref{eq:prop1}.

So \emph{no} violations of the property~\eqref{eq:prop1}
are found here. Equation~\eqref{inc1}, from
Ref.~\citep{brukneretal2001}, is simply incorrect, and the
seeming ``violations'' that arose from it disappear at
once if we write that equation more correctly as
\begin{equation}\label{eq:inc1clear}
\shb{\pr(\zBB\cond \zLone)} <\shb{\pr(\zBB\cond \zAA\land\zLL)},
\end{equation}
where we see that the left-hand side refers to the set-up $\zLone$,
whereas the right-hand side refers to the different set-up $\zLL$, so
that \emph{the comparison is between entropies relative to different
experiments}, and the equation \emph{does not concern
property~\eqref{eq:prop1}}. This fact escaped the attention of the
authors of Ref.~\citep{brukneretal2001}, partly because the
contexts were not explicitly written, and partly because the
probabilities were not explicitly calculated (so that the authors did
not notice that the probabilities for $\zBB$ had two numerically
different sets of values). The authors also seem to interpret the
expression ``$\shp{\zBB}$'' as implying that no observation must
precede $\zBB$, and the expression ``$\shp{\zBB\cond\zAA}$'' as
implying that the observation relative to $\zBB$ must be preceded by
the one relative to $\zAA$. As already remarked, this needs not be
the case.

It should be noted that the experimental arrangements
$\zLone$ and $\zLL$ are \emph{incompatible}, also in the
formal sense that their conjunction $\zLone\land\zLL$ is
false. We could erroneously see~$\zLL$ as a ``more
detailed'' description of $\zLone$, equivalent, for
example, to the conjunction of $\zLone$ \emph{and} the
proposition \prop{Moreover, a diagonal polarisation filter
is present between the photon source and the horizontal
filter}. But this is not the case: $\zLone$ states that
\emph{nothing} is present between the source and the
horizontal filter; if it had been otherwise, and $\zLone$
had left open the possibility that something unknown could
be between source and filter (a linear or circular
polarisation filter, or a mirror, or an opaque screen, or
something else), then we should have assigned a different
state to the photon reaching the horizontal filter, and
the calculation of the probabilities would have been very
different~\cite{fuchs2001}.

\subsection{Second quantum experiment}
\label{sec:second}
The second experiment~\cite[Fig.~5]{brukneretal2001} is as follows.
We send a spin-1/2 particle with spin up along the $z$ axis through a
Stern-Gerlach apparatus aligned along the axis $a$ that lies in the
$xz$ plane and forms an angle $\alpha$ with the $z$ axis. Let us
denote by $\zXX$ the set~$\set{\zXu,\zXd}$ with
$\zXu\defin{}$\prop{The particle comes out with spin up along $a$}
and $\zXd\defin{}$\prop{The particle comes out with spin down along
$a$}. We have the following probabilities, in the notation of
Ref.~\citep{brukneretal2001}:
\begin{gather}\label{eq:prx}
\bquoteq\pr(\zXu)=
\cos^2\tfrac{\alpha}{2},\quad\pr(\zXd)= \sin^2\tfrac{\alpha}{2}\equoteq,
\end{gather}
and a corresponding Shannon entropy which amounts to
\begin{align}\label{eq:shx}
\bquoteq\shp{\zXX}=
\shh{\cos^2\tfrac{\alpha}{2},\sin^2\tfrac{\alpha}{2}}\equoteq.
\end{align}
The particle then proceeds to a second Stern-Gerlach apparatus
aligned along the $x$ axis; let us denote the corresponding set of
alternatives by $\zYY\defin \set{\zYu,\zYd}$, where $\zYu$ and $\zYd$
are defined analogously to $\zXu$ and $\zXd$ above. The conditional
probabilities for $\zYY$ relative to the outcomes for $\zXX$ are:
\begin{gather}
\begin{aligned}
&\bquoteq\pr(\zYu\cond \zXu)=
\cos^2\bigl(\tfrac{\pu}{4}-\tfrac{\alpha}{2}\bigr)\equoteq,\\
&\bquoteq\pr(\zYd\cond \zXu)= \sin^2\bigl(\tfrac{\pu}{4}-\tfrac{\alpha}{2}\bigr)\equoteq, 
  \end{aligned}\label{eq:pryxu0}\\[1ex]
\begin{aligned}
&\bquoteq\pr(\zYu\cond \zXd)=\sin^2\bigl(\tfrac{\pu}{4}-\tfrac{\alpha}{2}\bigr)\equoteq,\\
&\bquoteq\pr(\zYd\cond \zXd)=\cos^2\bigl(\tfrac{\pu}{4}-\tfrac{\alpha}{2}\bigr)\equoteq,
\end{aligned}\label{eq:pryxd0}
\end{gather}
and we can easily calculate the following Shannon conditional
entropy:
\begin{equation}\label{eq:shyx}
\begin{aligned}
\bquoteq\shp{\zYY\cond \zXX}&=\cos^2\tfrac{\alpha}{2}\times
\shh{\cos^2\tfrac{\alpha}{2},\sin^2\tfrac{\alpha}{2}}
\\
&\quad {}+\sin^2\tfrac{\alpha}{2}\times
\shh{\sin^2\tfrac{\alpha}{2},\cos^2\tfrac{\alpha}{2}}\\
&=\shh{\cos^2\tfrac{\alpha}{2},\sin^2\tfrac{\alpha}{2}}\equoteq.
  \end{aligned}
\end{equation}
The sum of the entropies thus far calculated is
\begin{equation}\label{eq:sumyx}
\bquoteq\shp{\zXX}+\shp{\zYY\cond \zXX}= 2\, \shh{\cos^2\tfrac{\alpha}{2},\sin^2\tfrac{\alpha}{2}}\equoteq.
\end{equation}

Now we suppose to exchange the two Stern-Gerlach apparatus, putting
the one along $x$ before the one along $a$. We then find the
following probabilities for~$\zYY$:
\begin{gather}\label{eq:pry}
\bquoteq\pr(\zYu)=\tfrac{1}{2},\quad\pr(\zYd)=\tfrac{1}{2}\equoteq,
\end{gather}
with the associated Shannon entropy
\begin{align}\label{eq:shy}
\bquoteq\shp{\zYY}=\shh{\tfrac{1}{2},\tfrac{1}{2}}\equoteq.
\end{align}
The conditional probabilities, given by quantum theory, for
the set~$\zXX$ relative to the outcomes for $\zYY$ are:
\begin{gather}
\begin{aligned}
&\bquoteq\pr(\zXu\cond \zYu)=\sin^2\bigl(\tfrac{\pu}{4}-\tfrac{\alpha}{2}\bigr)\equoteq,\\
&\bquoteq\pr(\zXd\cond \zYu)=\cos^2\bigl(\tfrac{\pu}{4}-\tfrac{\alpha}{2}\bigr)\equoteq,
  \end{aligned}\label{eq:prxyu0}\\[1ex]
\begin{aligned}
&\bquoteq\pr(\zXu\cond \zYd)=\cos^2\bigr(\tfrac{\pu}{4}-\tfrac{\alpha}{2}\bigr)\equoteq,\\
&\bquoteq\pr(\zXd\cond \zYd)=\sin^2\bigr(\tfrac{\pu}{4}-\tfrac{\alpha}{2}\bigr)\equoteq,
  \end{aligned}\label{eq:prxyd0}
\end{gather}
and together with the probabilities~\eqref{eq:pry} they lead to the
conditional entropy
\begin{equation}\label{eq:shxy}
\begin{aligned}
\bquoteq\shp{\zXX\cond \zYY}&=\sin^2\tfrac{\alpha}{2}\times
\shh{\sin^2\tfrac{\alpha}{2},\cos^2\tfrac{\alpha}{2}}\\
&\quad{}+\cos^2\tfrac{\alpha}{2}\times
\shh{\cos^2\tfrac{\alpha}{2},\sin^2\tfrac{\alpha}{2}}\\
&=\shh{\cos^2\tfrac{\alpha}{2},\sin^2\tfrac{\alpha}{2}}\equoteq.
  \end{aligned}
\end{equation}
The sum of the entropies~\eqref{eq:shy} and~\eqref{eq:shxy} now yields
\begin{equation}\label{eq:sumxy}
\bquoteq\shp{\zYY}+\shp{\zXX\cond \zYY}
=\shh{\tfrac{1}{2},\tfrac{1}{2}}+\shh{\cos^2\tfrac{\alpha}{2},\sin^2\tfrac{\alpha}{2}}\equoteq,
\end{equation}
but this is in general (\eg\ for $\alpha=\pu/4$) different from the
sum~\eqref{eq:sumyx}, and thus we find that, in general,
\begin{equation}
\bquoteq\shp{\zXX} +\shp{\zYY\cond \zXX}\neq \shp{\zYY}+
\shp{\zXX\cond \zYY}\equoteq \quad\text{(seemingly)},\label{eq:inc2}
\end{equation}
in seeming contradiction with the property~\eqref{eq:prop2}.

However, we notice that two different relative positions of the
Stern-Gerlach apparatus were considered, in order to arrive at
Eqs.~\eqref{eq:sumyx} and~\eqref{eq:sumxy} respectively. This makes
the seeming contradiction above just an artifact produced, again, by the
comparison of Shannon entropies relative to two different
experimental arrangements, analogously to what happened in the
experiment with photons previously discussed. Also in this case, this
is shown by an explicit calculation of all the probabilities relative
to the experiments.

In the first set-up, which can be denoted by $\zM$, we
send the spin-1/2 particle to the Stern-Gerlach apparatus
oriented along $a$ (associated to the set of
alternatives~$\zXX$), which is in turn placed before the
one oriented along $x$ (associated to the set~$\zYY$).
Basic quantum-mechanical rules yield the following
probabilities (in our notation):
\begin{subequations}\label{eq:QMprob21}
\begin{gather}
\pr(\zXu\cond\zM)=\cos^2\tfrac{\alpha}{2},\quad
 \pr(\zXd\cond\zM)=\sin^2\tfrac{\alpha}{2},
\\[1ex]
\begin{aligned}
&\pr(\zYu\cond \zXu\land\zM)=\cos^2\bigl(\tfrac{\pu}{4}-\tfrac{\alpha}{2}\bigr),\\
&\pr(\zYd\cond
\zXu\land\zM)=\sin^2\bigl(\tfrac{\pu}{4}-\tfrac{\alpha}{2}\bigr),
\end{aligned}
\\[1ex]
\begin{aligned}
&\pr(\zYu\cond \zXd\land\zM)=\sin^2\bigl(\tfrac{\pu}{4}-\tfrac{\alpha}{2}\bigr),\\
&\pr(\zYd\cond
\zXd\land\zM)=\cos^2\bigl(\tfrac{\pu}{4}-\tfrac{\alpha}{2}\bigr),
\end{aligned}
\end{gather}
\end{subequations}
and these are sufficient to calculate, by the product
rule~\eqref{eq:productrule}, the joint probabilities
\begin{subequations}\label{eq:jointprex21}
\begin{align}
\zauu&\defin\pr(\zXu\land \zYu\cond\zM)=
\cos^2\tfrac{\alpha}{2}\cdot\cos^2\bigl(\tfrac{\pu}{4}-\tfrac{\alpha}{2}\bigr),\\
\zaud&\defin\pr(\zXu\land \zYd\cond\zM)=
\cos^2\tfrac{\alpha}{2}\cdot\sin^2\bigl(\tfrac{\pu}{4}-\tfrac{\alpha}{2}\bigr),\\
\zadu&\defin\pr(\zXd\land \zYu\cond\zM)=
\sin^2\tfrac{\alpha}{2}\cdot\sin^2\bigl(\tfrac{\pu}{4}-\tfrac{\alpha}{2}\bigr),\\
\zadd&\defin\pr(\zXd\land \zYd\cond\zM)=
\sin^2\tfrac{\alpha}{2}\cdot\cos^2\bigl(\tfrac{\pu}{4}-\tfrac{\alpha}{2}\bigr).
\end{align}
\end{subequations}

The formulae above show clearly that, as remarked in
\sect\ref{sec:notat}, it \emph{does} make sense to consider the
\emph{composite probability} of the outcomes of the two
\emph{temporally separated} observations, as a composite probability
needs have nothing to do with the fact that the observations are
performed ``simultaneously'' or not.

From the joint probabilities, we can compute all
probabilities involved in this set-up. Applying the marginal
probability rule~\eqref{eq:marginalprob} and Bayes'
rule~\eqref{eq:bayesrule} we find:
\begin{gather}
\pr(\zXu\cond \zM)=\zauu+\zaud,\quad
\pr(\zXd\cond \zM)=\zadu+\zadd,
\label{eq:prx1}
\\[1ex]    
\pr(\zYu\cond \zM)=\zauu+\zadu,\quad
\pr(\zYd\cond \zM)=\zaud+\zadd,
\label{eq:pry1}\\[1ex]
\begin{aligned}
&\pr(\zXu\cond \zYu\land\zM)=\frac{\zauu}{\zauu+\zadu},\\
&\pr(\zXd\cond \zYu\land\zM)=\frac{\zadu}{\zauu+\zadu},
\end{aligned}\label{eq:prxyu}\\[1ex]
\begin{aligned}
&\pr(\zXu\cond \zYd\land\zM)=\frac{\zaud}{\zaud+\zadd},\\
&\pr(\zXd\cond \zYd\land\zM)=\frac{\zadd}{\zaud+\zadd},
\end{aligned}\label{eq:prxyd}\\[1ex]
\begin{aligned}
&\pr(\zYu\cond \zXu\land\zM)=\frac{\zauu}{\zauu+\zaud},\\
&\pr(\zYd\cond \zXu\land\zM)=\frac{\zaud}{\zauu+\zaud},
\end{aligned}\label{eq:pryxu}\\[1ex]
\begin{aligned}
&\pr(\zYu\cond \zXd\land\zM)=\frac{\zadu}{\zadu+\zadd},\\
&\pr(\zYd\cond \zXd\land\zM)=\frac{\zadd}{\zadu+\zadd}.
\end{aligned}\label{eq:pryxd}
\end{gather}
Here the probabilities~\eqref{eq:QMprob21} have been re-written
in terms of the joint probabilities.

A look at Eqs.~\eqref{eq:prxyu} and~\eqref{eq:prxyd} shows that
Bayes' rule also applies in ``quantum experiments'', and provides a
counter-example to the incorrect statement that the expression `$\zXu\cond
\zYu$' would mean ``$\zYu$ precedes $\zXu$''.

We can proceed to calculate the Shannon entropies
\begin{widetext}
\begin{align}
&\shb{\pr(\zXX\cond \zM)}= -\shk[(\zauu+\zaud) \ln(\zauu+\zaud)
+ (\zadu+\zadd) \ln (\zadu+\zadd)],\label{eq:shx1}\\
&\shb{\pr(\zYY\cond \zM)}= -\shk[(\zauu+\zadu) \ln(\zauu+\zadu) 
+(\zaud+\zadd) \ln (\zaud+\zadd)],\label{eq:shy1}
\end{align}
as well as the conditional entropies
\begin{align}
\begin{split}
\shb{\pr(\zYY\cond \zXX\land\zM)}&=\shk\,(\zauu+ \zaud)\,
\biggl( -\frac{\zauu}{\zauu+\zaud} \ln\frac{\zauu}{\zauu+\zaud}
-\frac{\zaud}{\zauu+\zaud} \ln\frac{\zaud}{\zauu+\zaud}\biggr)\\*
&\quad {}+ \shk\,(\zadu + \zadd)\, \biggl(-\frac{\zadu}{\zadu+\zadd}
\ln\frac{\zadu}{\zadu+\zadd} -\frac{\zadd}{\zadu+\zadd}
\ln\frac{\zadd}{\zadu+\zadd}
\biggr)\\
&=\shk\,[-\zauu\ln\zauu -\zadu\ln\zadu -\zaud\ln\zaud
-\zadd\ln\zadd\\*
&\qquad\;{}+(\zauu+\zaud) \ln(\zauu+\zaud) + (\zadu+\zadd) \ln
(\zadu+\zadd)],
\end{split}\label{eq:shyx1}\\
\begin{split} \shb{\pr(\zXX\cond \zYY\land\zM)}&= \shk\,(\zauu+ \zadu)\,
\biggl( -\frac{\zauu}{\zauu+\zadu} \ln\frac{\zauu}{\zauu+\zadu}
-\frac{\zadu}{\zauu+\zadu} \ln\frac{\zadu}{\zauu+\zadu}\biggr)\\
&\quad {}+ \shk\,(\zaud + \zadd)\, \biggl(-\frac{\zaud}{\zaud+\zadd}
\ln\frac{\zaud}{\zaud+\zadd} -\frac{\zadd}{\zaud+\zadd}
\ln\frac{\zadd}{\zaud+\zadd}
\biggr)\\
&=\shk\,[- \zauu\ln\zauu -\zadu\ln\zadu -\zaud\ln\zaud
-\zadd\ln\zadd\\
&\qquad\;+(\zauu+\zadu) \ln(\zauu+\zadu) + (\zaud+\zadd) \ln
(\zaud+\zadd)],
\end{split}\label{eq:shxy1}
\end{align}
where the expressions have been simplified making use of the
additivity property of the logarithm.

Finally, from Eqs.~\eqref{eq:shx1} and~\eqref{eq:shyx1},
\eqref{eq:shy1} and~\eqref{eq:shxy1}, we find
\begin{equation}\label{eq:confpr2}
\begin{aligned}
\shb{\pr(\zXX\cond \zM)}+ \shb{\pr(\zYY\cond \zXX\land\zM)}&\equiv
\shb{\pr(\zYY\cond \zM)} + \shb{\pr(\zXX\cond \zYY\land\zM)}\\
&= - \shk\,(\zauu\ln\zauu +\zadu\ln\zadu + \zaud\ln\zaud +
\zadd\ln\zadd)\\
&\equiv\shb{\pr(\zAB\cond\zM)},
\end{aligned}
\end{equation}
\end{widetext}
whence \emph{we see that the property~\eqref{eq:prop2} is satisfied}
(we find, \eg, that $\shb{\pr(\zXX\cond \zM)}+ \shb{\pr(\zYY\cond
\zXX\land\zM)} \equiv\shb{\pr(\zYY\cond \zM)} + \shb{\pr(\zXX\cond
\zYY\land\zM)}\equiv\shb{\pr(\zAB\cond\zM)}\approx 1.20\bit$, for
$\alpha=\pu/4$).

We note that the way in which Eq.~\eqref{eq:confpr2} has been found
does not depend on the numerical values of the probabilities
$\set{\za_{ij}}$, but only on the additivity property of the
logarithm; so the calculations above can be seen as a mathematical
proof of the property~\eqref{eq:prop2} for the special case of sets
with only two alternatives.

If we invert the positions of the two Stern-Gerlach apparatus,
placing the one oriented along $x$ (associated to the set $\zYY$) before the
one oriented along $a$ (associated to the set $\zXX$), we then realise a
new, \emph{different} experimental arrangement, which can be denoted by
$\zML$. The probabilities for the sets of alternatives $\zXX$ and
$\zYY$ will thus \emph{differ} from those in $\zM$: we have indeed
\begin{subequations}\label{eq:QMprob22}
\begin{gather}
\pr(\zYu\cond\zML)=\tfrac{1}{2},\quad\pr(\zYd\cond\zML)=\tfrac{1}{2},\\[1ex]
\begin{aligned}
&\pr(\zXu\cond \zYu\land\zML)=\sin^2\bigl(\tfrac{\pu}{4}-\tfrac{\alpha}{2}\bigr),\\
&\pr(\zXd\cond
\zYu\land\zML)=\cos^2\bigl(\tfrac{\pu}{4}-\tfrac{\alpha}{2}\bigr),
\end{aligned}\\[1ex]
\begin{aligned}
&\pr(\zXu\cond \zYd\land\zML)=\cos^2\bigr(\tfrac{\pu}{4}-\tfrac{\alpha}{2}\bigr),\\
&\pr(\zXd\cond
\zYd\land\zML)=\sin^2\bigr(\tfrac{\pu}{4}-\tfrac{\alpha}{2}\bigr).
\end{aligned}
\end{gather}
\end{subequations}

It is clear that, from this point on, we can proceed as in the
analysis of the first set-up, obtaining
\begin{multline}
\shb{\pr(\zXX\cond \zML)}+ \shb{\pr(\zYY\cond
\zXX\land\zML)}\\
\begin{aligned}
&=\shb{\pr(\zYY\cond \zML)} + \shb{\pr(\zXX\cond \zYY\land\zML)}\\
&\begin{aligned}
{}=-\shk\,(&\zauu'\ln\zauu' +\zadu'\ln\zadu'\\
&{}+ \zaud'\ln\zaud' + \zadd'\ln\zadd'),
\end{aligned}\\
&\equiv\shb{\pr(\zAB\cond\zML)}
\end{aligned}
\end{multline}
where the $\set{\za_{ij}'}$ are the values of the joint probabilities
$\set{\pr(\zX_i\land\zY_j\cond\zML)}$, different, in general, from
the $\set{\za_{ij}}$. In any case, \emph{property~\eqref{eq:prop2} is
again satisfied in the new context} (we have, \eg,
$\shb{\pr(\zXX\cond \zML)}+ \shb{\pr(\zYY\cond
\zXX\land\zML)}\equiv\shb{\pr(\zYY\cond \zML)} + \shb{\pr(\zXX\cond
\zYY\land\zML)}\equiv\shb{\pr(\zAB\cond\zML)}\approx 1.60\bit$, for
$\alpha=\pu/4$).

Thus we have found \emph{no} inconsistencies in this
second experiment either: Equation~\eqref{eq:inc2}, from
Ref.~\citep{brukneretal2001}, is simply incorrect, and can
more correctly be written as
\begin{multline}\label{eq:inc2clear}
\shb{\pr(\zXX\cond \zM)}+  \shb{\pr(\zYY\cond\zXX\land\zM)}\\
{}\neq\shb{\pr(\zYY\cond \zML)}+\shb{\pr(\zXX\cond\zYY\land\zML)},
\end{multline}
or equivalently and more briefly as
\begin{equation}\label{eq:ex2sumclear}
\shb{\pr(\zXX\land\zYY\cond\zM)}\neq\shb{\pr(\zXX\land\zYY\cond\zML)},
\end{equation}
and its content is that the Shannon entropies in the
experiment $\zM$ are in general different from those in
the different experiment $\zML$. This does not surprise
us, since the experiments have different set-ups.

The fact that Eq.~\eqref{eq:inc2} \emph{does not pertain to
property~\eqref{eq:prop2}} is not noticed in
Ref.~\citep{brukneretal2001}, again because the contexts are not
kept explicit in the notation. Moreover, the expression
``$\shp{\zYY\cond \zXX}$'' is considered there as implying that the
observation corresponding to $\zXX$ is performed \emph{before} the
one corresponding to $\zYY$:\footnote{\Cf\ the discussion of the
formula ``$\shp{\zXX}+\shp{\zYY\cond \zXX} = \shp{\zYY}+\shp{\zXX\cond
\zYY}$'' in Ref.~\citep[\sect{}III]{brukneretal2001}.} it is for
this reason that, in order to calculate ``$\shp{\zYY\cond \zXX}$'', the
authors consider the experiment in which the observation
corresponding to $\zXX$ is performed before the one corresponding to
$\zYY$, but then, in order to calculate ``$\shp{\zXX\cond \zYY}$'', they
feel compelled to change the order of the observations --- with the
only effect of changing the whole problem and all probabilities
instead! But, as already remarked, the conditional symbol `$\cond$'
does not have that meaning. The point is that the temporal order of
acquisition of knowledge about two physical events does not
necessarily correspond to the temporal order in which these events
occur.

\subsection{Classical experiment}
\label{sec:third}
Together with the two quantum experiments, the authors of
Ref.~\citep{brukneretal2001} also present an example of a
classical experiment in which, they claim, the
property~\eqref{eq:prop2} is not violated. It is useful to
re-analyse this example as well, in order to show that, in
fact, it is \emph{not} an instance of confirmation of the
property~\eqref{eq:prop2}, because it, too, involves two
different experimental arrangements.

The idea~\cite[Fig.~3]{brukneretal2001} is as follows. We fill a box
with four balls of different colours (black and white) and
compositions (plastic and wood). There are two black plastic balls,
one white plastic ball, one white wooden ball. We shake the box, draw
a ball blindfold, and consider the set $\zXX\defin\set{\zXb,\zXw}$
for the ball's being black or white. If the ball is black, then we
put all black balls in a new box, draw a new ball from this box, and
consider the set $\zYY\defin\set{\zYp,\zYw}$ for this ball's being
plastic or wooden. We proceed analogously if the first drawn ball
was white instead.\footnote{In Brukner and Zeilinger's original example, the
black and white balls are put into two separate boxes after the first
draw, and a ball is drawn \emph{from each box} separately. 
But these two final draws are then related to two separate
observations, not one, so that in total we have \emph{three}
observations in this experiment, and the formula~\eqref{eq:prop2} is
not even appropriate. The experiment has thus been modified here, in
order to preserve the two authors' original intention.} Let us denote the
set-up just described by $\zN$.

The probabilities of first drawing a black or a white ball are
respectively $\pr(\zXb\cond\zN)=1/2$ and $\pr(\zXw\cond\zN)=1/2$
and thus their Shannon entropy is
\begin{equation}
  \label{eq:shcolour1}
  \shb{\pr(\zXX\cond\zN)}=\shh{\tfrac{1}{2},\tfrac{1}{2}}=1\bit.
\end{equation}
The conditional probabilities of the second drawn ball's being
plastic or wooden, given the outcome of the first observation, are
\begin{gather}
\pr(\zYp\cond\zXb\land\zN)=1,\quad\pr(\zYw\cond\zXb\land\zN)=0,\\
\intertext{if the first result was `black', and} 
\pr(\zYp\cond\zXw\land\zN)=\tfrac{1}{2},\quad\pr(\zYw\cond\zXw\land\zN)=\tfrac{1}{2},
\end{gather}
if it was `white'. From these probabilities the following Shannon
conditional entropy can be computed:
\begin{equation}
\label{eq:shcomp1}
\begin{aligned}
\shb{\pr(\zYY\cond\zXX\land\zN)}&=\tfrac{1}{2}\,\shp{1,0}+
\tfrac{1}{2}\,\shh{\tfrac{1}{2},\tfrac{1}{2}}\\
&=\tfrac{1}{2}\mult 0\bit+\tfrac{1}{2}\mult 1\bit
=0.5\bit.
\end{aligned}
\end{equation}
Combining Eqs.~\eqref{eq:shcolour1} and~\eqref{eq:shcomp1} we obtain
\begin{equation}
\label{eq:shsum1}
\shb{\pr(\zXX\cond\zN)}+\shb{\pr(\zYY\cond\zXX\land\zN)}=1.5\bit.
\end{equation}

Now we suppose to make the observations in inverse order instead. We
shake the initial box, draw a ball, consider first the set
$\zYY\defin\set{\zYp,\zYw}$ for the ball's being plastic or wooden.
Depending on the outcome we fill a new box either with the plastic or
the wooden balls, and draw a new ball; then we consider the set
$\zXX\defin\set{\zXb,\zXw}$ for the new ball. Let us denote this new
experiment by $\zNL$. It is clear that $\zN$ and $\zNL$ are really
different experiments, for the following reason: in $\zN$, between
the two draws, the second box contains black or white balls; while in
$\zNL$ it contains plastic or wooden balls.

The probabilities for the set $\zYY$ this time are
$\pr(\zYp\cond\zNL)=\tfrac{3}{4}$ and
$\pr(\zYw\cond\zNL)=\tfrac{1}{4}$, with an entropy
\begin{equation}\label{eq:shcomp2}
\shb{\pr(\zYY\cond\zNL)}=\shh{\tfrac{3}{4},\tfrac{1}{4}}\approx0.81\bit,
\end{equation}
while the conditional probabilities for $\zXX$ are
\begin{gather}
\begin{aligned}
&\pr(\zXb\cond\zYp\land\zNL)=\tfrac{2}{3},\\
&\pr(\zXw\cond\zYp\land\zNL)=\tfrac{1}{3},
\end{aligned}\\
\intertext{if the first result was `plastic', and}
\begin{aligned}
&\pr(\zXb\cond\zYw\land\zNL)=0,\\
&\pr(\zXw\cond\zYw\land\zNL)=1,
\end{aligned}
\end{gather}
if it was `wooden'. The conditional entropy is
\begin{equation}
\label{eq:shcol2}
\begin{aligned}
\shb{\pr(\zXX\cond\zYY\land\zNL)}&=\tfrac{3}{4}\,
\shh{\tfrac{2}{3},\tfrac{1}{3}}+\tfrac{1}{4}\,\shp{0,1}\\
&\approx\tfrac{3}{4}\mult 0.92\bit+\tfrac{1}{3}\mult 0\bit\\
&\approx0.69\bit,
\end{aligned}
\end{equation}
and adding this time Eqs.~\eqref{eq:shcomp2} and~\eqref{eq:shcol2}, we
find
\begin{equation}
\label{eq:shsum2}
\shb{\pr(\zYY\cond\zNL)}+\shb{\pr(\zXX\cond\zYY\land\zNL)}=1.5\bit.
\end{equation}

We see that the entropies~\eqref{eq:shsum1} and \eqref{eq:shsum2} are equal,
\begin{multline}
\shb{\pr(\zXX\cond\zNL)}+\shb{\pr(\zYY\cond\zXX\land\zNL)} \\
{}=\shb{\pr(\zYY\cond\zNL)}+
\shb{\pr(\zXX\cond\zYY\land\zNL)},\label{eq:eqclas}
\end{multline}
which is equivalent to
\begin{gather}
\shb{\pr(\zXX\land\zYY\cond\zN)}=
\shb{\pr(\zXX\land\zYY\cond\zNL)},\label{eq:eqclascorsimp}
\end{gather}
but it is clear that \emph{this is not the statement of
property~\eqref{eq:prop2}}, because the right- and left-hand
sides of this equation refer to two different experiments. The
content of the equality above is only that the Shannon entropies for
the probability distributions for the composite set of alternatives
$\zXX\land\zYY$ are equal in the two experiments $\zN$ and~$\zNL$.

\section{\label{sec:fourth}The r\^ole of ``quantumness''}

We have shown thus far that the quantum experiments in
Ref.~\citep{brukneretal2001} did \emph{not} involve any violation of
the Shannon entropy's properties. However, we may still imagine
someone raising the following argument:

\begin{quotation}
``Very well, the properties~\eqref{eq:prop} are not
violated in any experiment. But one notices that the
equality \emph{for different contexts}
\begin{equation}\tag*{(\ref{eq:eqclascorsimp})$_\text{r}$}
\shb{\pr(\zXX\land\zYY\cond\zN)}=\shb{\pr(\zXX\land\zYY\cond\zNL)},
\end{equation}
while holding in the classical experiment of
\sect\ref{sec:third}, does not hold in general in the
quantum one of \sect\ref{sec:second}, where it is found
instead that
\begin{equation}\tag*{(\ref{eq:ex2sumclear})$_\text{r}$}
\shb{\pr(\zXX\land\zYY\cond\zM)}\neq\shb{\pr(\zXX\land\zYY\cond\zML)}.
\end{equation}
From this particular case, one can see that in classical
experiments the Shannon entropy remains the same if the
temporal order of observations is changed, whereas in
quantum experiments the entropy changes together with the
change in temporal order. This phenomenon is thus a
peculiarity of the quantum nature of the experiments --- a
sort of `quantum-context-dependence' of the Shannon
entropy.''
\end{quotation}

But this argument has no validity, of course. The Shannon
entropy is \emph{always} ``context dependent'', and this
comes from the fact that probabilities are \emph{always}
context dependent, in \emph{both} classical and quantum
experiments. We can further and illustrate this fact by
means of two more experiments, which will also serve as
counter-examples of the ones already discussed.

\subsection{\label{sec:secondco}First counter-example}

The first counter-example is a modification, based on the examples
presented by Kirkpatrick~\cite{kirkpatrick2001,kirkpatrick2002} of
the experiment with the balls discussed in
\sect\ref{sec:third}.\footnote{Cf.\ also the toy models discussed by
Hardy~\cite{Hardy1999} and Spekkens~\citep{Spekkens2004}.} The balls
are in addition big or small as well now: there are one big black
plastic ball, one small black plastic ball, one small white plastic
ball, and one small white wooden ball.

Initially, we prepare the box so that it contains only all small
balls. Then we shake the box, draw a ball blindfold, and consider the
set $\zXX\defin\set{\zXb,\zXw}$ for the ball's being black or white.
If the drawn ball is black, then we prepare the box so that it
contains only \emph{all} black balls (also the big black one that was
initially not in the box), draw a new ball from this box, and
consider the set $\zYY\defin\set{\zYp,\zYw}$ for this ball's being
plastic or wooden. We proceed analogously if the first drawn ball was
white.\footnote{The Reader should not, too simply, identify the
``system'' here with some specific group of balls. It is rather
associated with a variable collection of balls, in analogy with a
classical \emph{open} system associated with variable number (and
species) of particles.} It is easy to see that, in the set-up just
described, denoted by $\zK$, we have the following probabilities:
\begin{gather}
\pr(\zXb\cond\zK)=\tfrac{1}{3},\quad\pr(\zXw\cond\zK)=\tfrac{2}{3},\\  
\pr(\zYp\cond\zXb\land\zK)=1,\quad\pr(\zYw\cond\zXb\land\zK)=0,\\
\pr(\zYp\cond\zXw\land\zK)=\tfrac{1}{2},\quad\pr(\zYw\cond\zXw\land\zK)=\tfrac{1}{2}.
\end{gather}
The Shannon entropies are
\begin{gather}
\shb{\pr(\zXX\cond\zK)}=
\shh{\tfrac{1}{3},\tfrac{2}{3}}\approx0.92\bit,\\[1ex]
\begin{aligned}
 \shb{\pr(\zYY\cond\zXX\land\zK)}
&=\tfrac{1}{3}\,\shp{1,0}+\tfrac{2}{3}\,\shh{\tfrac{1}{2},\tfrac{1}{2}}\\
&=\tfrac{1}{3}\mult 0\bit+\tfrac{2}{3}\mult 1\bit 
\approx0.67\bit,
\end{aligned}
\end{gather}
and their sum is
\begin{multline}\label{eq:sumclnew1}
\shb{\pr(\zXX\cond\zK)}+\shb{\pr(\zYY\cond\zXX\land\zK)}\\
\begin{aligned}
&\equiv\shb{\pr(\zYY\cond\zK)}+\shb{\pr(\zXX\cond\zYY\land\zK)}\\
&\equiv\shb{\pr(\zXX\land\zYY\cond\zK)} \approx1.58\bit.
\end{aligned}
\end{multline}

Now let us consider the set-up $\zKL$, in which the observations are
made in reverse order, but with the same general procedure. The
probabilities are then:
\begin{gather}
\pr(\zYp\cond\zKL)=\tfrac{2}{3},\quad\pr(\zYw\cond\zKL)=\tfrac{1}{3},\\[1ex]  
\begin{align}
&\pr(\zXb\cond\zYp\land\zKL)=\tfrac{2}{3},\\
&\pr(\zXw\cond\zYp\land\zKL)=\tfrac{1}{3},
\end{align}\\[1ex]
\begin{align}
&\pr(\zXb\cond\zYw\land\zKL)=0,\\
&\pr(\zXw\cond\zYw\land\zKL)=1.
\end{align}
\end{gather}
These lead to the entropies
\begin{gather}
\shb{\pr(\zYY\cond\zKL)}=\shh{\tfrac{2}{3},\tfrac{1}{3}}\approx
0.92\bit,\\[1ex]
\begin{aligned}
\shb{\pr(\zXX\cond\zYY\land\zKL)}
&=\tfrac{2}{3}\,\shh{\tfrac{2}{3},\tfrac{1}{3}}+
\tfrac{1}{3}\,\shp{0,1}\\
&\approx\tfrac{2}{3}\mult 0.92\bit+
\tfrac{1}{3}\mult 0\bit\\
&\approx0.61\bit,
\end{aligned}
\end{gather}
and the sum
\begin{multline}\label{eq:sumclnew2}
\shb{\pr(\zYY\cond\zKL)}+
\shb{\pr(\zXX\cond\zYY\land\zKL)}\\
\begin{aligned}
&\equiv\shb{\pr(\zXX\cond\zKL)}+\shb{\pr(\zYY\cond\zXX\land\zKL)}\\
&\equiv \shb{\pr(\zXX\land\zYY\cond\zKL)} \approx1.53\bit.
\end{aligned}
\end{multline}

Comparing Eqs.~\eqref{eq:sumclnew1} and \eqref{eq:sumclnew2} we find
\begin{equation}
\shb{\pr(\zXX\land\zYY\cond \zK)}\neq\shb{\pr(\zXX\land\zYY\cond \zKL)}.
\end{equation}
Hence, for this experiment, of a clearly classical nature, we obtain
different statistics and entropies depending on the order in which we
observe colour and composition.\footnote{We may note, incidentally,
that it has long been known that the \emph{thermodynamic} entropy
of a classical thermodynamic system in a non-equilibrium state
depends on its complete previous history~\cite{truesdell1969}.}

Note, in any case, that in each of the two set-ups ---
Eqs.~\eqref{eq:sumclnew1} and~\eqref{eq:sumclnew2} --- the
properties~\eqref{eq:prop} are always satisfied, just as in the
quantum experiments.

\subsection{\label{sec:thirdco}Second counter-example}

The second counter-example is of a
quantum-me\-chan\-i\-cal nature. It runs precisely like
the experiment with spin-1/2 particles discussed in
\sect\ref{sec:second}, except that now the particle has
initially spin up, not along the $z$ axis, but along the
axis $b$ that bisects the angle $\widehat{ax}$ (\ie, $b$
lies in the $xaz$ plane and forms an angle
$\beta\defin\pu/4-\alpha/2$ with both the $x$ and $a$
axes; remember that $\alpha$ is the angle $\widehat{az}$).
The analysis of this experiment proceeds completely along
the lines of the re-analysis of \sect\ref{sec:second}, if
we introduce the contexts $\zMM$ and $\zMML$, and change
Eqs.~\eqref{eq:QMprob21} with
\begin{subequations}\label{eq:QMprobnew1}
\begin{gather}
\pr(\zXu\cond\zMM)=\cos^2\tfrac{\beta}{2},\quad
\pr(\zXd\cond\zMM)=\sin^2\tfrac{\beta}{2},
\\[1ex]
\begin{aligned}
    &\pr(\zYu\cond \zXu\land\zMM)=\cos^2\beta,\\
    &\pr(\zYd\cond
    \zXu\land\zMM)=\sin^2\beta,
  \end{aligned}\\[1ex]
\begin{aligned}
    &\pr(\zYu\cond \zXd\land\zMM)=\sin^2\beta,\\
    &\pr(\zYd\cond
    \zXd\land\zMM)=\cos^2\beta,
  \end{aligned}
\end{gather}
\end{subequations}
and Eqs.~\eqref{eq:QMprob22} with
\begin{subequations}\label{eq:QMprobnew2}
\begin{gather}
\pr(\zYu\cond\zMML)=\cos^2\tfrac{\beta}{2},\quad
\pr(\zYd\cond\zMML)=\sin^2\tfrac{\beta}{2},
\\[1ex]
\begin{aligned}
    &\pr(\zXu\cond \zYu\land\zMML)=\sin^2\beta,\\
    &\pr(\zXd\cond
    \zYu\land\zMML)=\cos^2\beta,
  \end{aligned}\\[1ex]
\begin{aligned}
    &\pr(\zXu\cond \zYd\land\zMML)=\cos^2\beta,\\
    &\pr(\zXd\cond
    \zYd\land\zMML)=\sin^2\beta.
  \end{aligned}
\end{gather}
\end{subequations}
It is obvious that this leads to the equalities
$\pr(\zX_i\land\zY_j\cond\zMM) = \pr(\zX_i\land\zY_j\cond\zMML)$
and eventually to the equality\begin{equation}\label{eq:counter1}
\shb{\pr(\zXX\land\zYY\cond \zMM)}=\shb{\pr(\zXX\land\zYY\cond\zMML)},
\end{equation}
exactly as it happened in the experiment with the balls of
\sect\ref{sec:third} (for $\alpha=\pu/4$, \eg, we have
$\shb{\pr(\zXX\land\zYY\cond \zMM)}\approx 0.83\bit$). But
here the experiment is a purely quantum one: we see indeed
that the observables \emph{do not commute} here, the
initial state is \emph{pure}, and its density matrix is
\emph{not diagonal} in either of the observables' bases.
Compare this result with the discussion in
Ref.~\citep{brukneretal2000}.

\section{Conclusions\label{concl}}
We have shown that the properties of the Shannon entropy
are not violated in quantum experiments, contrary to the
conclusions of Ref.~\citep{brukneretal2001}. In that
paper, an idiosyncratic temporal interpretation of the
conditional symbol `$\cond$' and of the conjunction symbol
`$\land$' leads the authors to change experimental set-ups
in order to calculate various entropies. As a result, they
compare entropies relative to different experimental
arrangements instead, and do not notice that the
probabilities are also different, and so their results
(Eqs.~\eqref{inc1} and~\eqref{eq:inc2} in this paper),
besides not being formally correctly written, do not
pertain to the properties of the Shannon entropy
(Eqs.~\eqref{eq:prop}), which refer to a single,
well-defined experiment and hold unconditionally.

The peculiar results arising from the comparison of entropies
relative to different experimental contexts can thus appear or not
appear in any kind of experiments, classical as well as quantum; this
has been shown by means of two counter-examples: a classical one, in
which a change in the temporal order of the experiment leads to a
change in entropy values, and a genuinely quantum one, in which the
same temporal change leads to no entropy changes.

A conclusion is that Bohr's dictum ought to be observed also in
mathematical notation, and not only in the analysis of quantum
phenomena, but in the analysis of classical phenomena as well,
because probabilities --- and, consequently, Shannon entropies ---
always depend on ``the \emph{whole experimental
arrangement}''~\cite{bohr1949} taken into account.

\begin{acknowledgements}
I thank Professor Gunnar Bj\"ork for continuous
encouragement and advice, Professor Sandra Brunsberg for
advice, and Professor Kim Kirkpatrick for useful
discussions. Financial support from the Foundation Angelo
Della Riccia and the Foundation BLANCEFLOR
Boncompagni-Ludovisi n\'ee Bildt is gratefully
acknowledged.
\end{acknowledgements}

\providecommand{\href}[2]{#2}
\providecommand{\eprint}[2][http://arxiv.org/abs/]{\href{#1#2}{\texttt{#2}}}
\newcommand{\arxiveprint}{
\eprint}
\newcommand{\citein}[1]{\textnormal{\citet{#1}}}

\end{document}